# Coherent Phonon Transport in Two-dimensional Graphene Superstructures

Usama Choudhry, Shengying Yue and Bolin Liao[*]

Department of Mechanical Engineering, University of California, Santa Barbara, CA 93106 USA

**Abstract**

Coherent wave effects of thermal phonons hold promise of transformative opportunities in thermal transport control but remain largely unexplored due to the small wavelength of thermal phonons, typically below a few nanometers. This small length scale indicates that, instead of artificial phononic crystals, a more promising direction is to examine the coherent phonon effects in natural materials with hierarchical superstructures matching the thermal phonon wavelength. In this work, we use first-principles simulations to characterize the previously unstudied thermal properties of dodeca-graphene (D-graphene) and tetra-graphene (T-graphene), two-dimensional carbon allotropes based upon the traditional graphene structure but containing a secondary, in-plane periodicity. We find that despite very similar atomic structure and bonding strength, D-graphene and T-graphene possess significantly different thermal properties than that of pristine graphene. At room temperature, the calculated thermal conductivity of D-graphene and T-graphene is 600 $Wm^{-1}K^{-1}$ and 800 $Wm^{-1}K^{-1}$ compared to over 3000 $Wm^{-1}K^{-1}$ for graphene. We attribute these distinct properties to the presence of naturally occurring, low frequency optical phonon modes that display characteristics of phonon coherence and arise from a folding of the acoustic modes and the associated frequency gap opening, a phenomenon also found in superlattices where an out of plane periodicity is introduced. Furthermore, we observe significantly enhanced Umklapp scatterings in D- and T-graphene that largely suppress the hydrodynamic phonon transport in pristine graphene. Our study presents D-graphene and T-graphene as ideal model systems to explore the coherent phonon effects in 2D and demonstrates the potential of using coherent phonon effects to significantly modify thermal transport of 2D materials without making drastic changes to their fundamental compositions.

[*] To whom correspondence should be addressed. Electronic mail: bliao@ucsb.edu

**Introduction**

The *in situ* control of wave propagation through conventional media is of great interest and significance to many problems across numerous fields. Significant research has been put into the design and engineering of metamaterials, artificial structures that typically have periodic orders matching the wavelength of the propagating waves (e.g. light and sound), resulting in highly unusual bulk material properties that are typically not found naturally. Photonic metamaterials, or photonic crystals, have provided unprecedented capability to control the propagation of light[1]. Acoustic metamaterials have been designed with negative indices of refraction[2] or negative bulk moduli[3], for uses in applications such as acoustic cloaking or acoustic imaging. Successful manipulation of wave propagation in materials typically requires a feature size comparable to or smaller than the relevant wavelengths present. This makes the design of effective thermal metamaterials difficult, as the wavelength of thermally relevant phonons is typically on the order of a few nanometers. Coherent wave effects of thermal phonons hold promise of transformative opportunities in thermal transport control[4,5], but remain largely unexplored due to the small wavelength of thermal phonons. Luckyanova *et al.* observed enhanced thermal conductivity in GaAs/AlAs superlattices[6] with a 24 nm period due to coherent transport of long wavelength acoustic phonons across the interfaces, and further discovered coherent backscattering, or localization, of phonons in the same structure[7] with ErAs dots randomly distributed at the interfaces. Ravichandran *et al.* observed a transition from incoherent phonon scattering at interfaces to coherent phonon transport in epitaxially grown $SrTiO_3/CaTiO_3$ and $SrTiO_3/BaTiO_3$ oxide superlattices[8] when the period of the superlattices becomes approximately 1 nm. These seminal studies exemplify the great potential of tuning phonon thermal conductivity via coherent interference of thermal phonons, but were conducted only at cryogenic temperature so that the characteristic size of the artificial structures matches the wavelength of the phonons that dominate the thermal transport. As pointed out by Lee et al., phonon coherence becomes negligible in silicon phononic crystals with a periodicity above 100 nm when the temperature is above 14 K[9], where incoherent phonon scatterings at the boundaries dominate the thermal transport[10]. Computationally, the coherent hybridization of propagating long-wavelength acoustic phonons with local resonances introduced by artificial periodic structures has been shown to significantly alter the thermal conductivity, whereas the required size of the artificial structures is typically a few nanometers[11,12]. This small length scale indicates that, instead of artificial phononic crystals, a more promising route to observing and exploiting coherent phonon effects at a higher temperature is to examine the coherent phonon effects in natural materials with hierarchical superstructures matching the thermal phonon wavelength. Recently, Guo *et al.* used first-principles simulations to examine the phonon transport across van der Waals heterostructures with interlayer distances matching the thermal phonon wavelength, where they observed strong effects of phonon coherence leading to highly anisotropic thermal conduction and ultralow thermal conductivity below the amorphous limit[13]. Due to the weak interlayer van der Waals interaction, however, the cross-plane thermal conductivity of van der Waals heterostructures is intrinsically low, which limits the achievable tuning range via phonon coherence. In this work, we use first-principles simulations to investigate in-plane phonon transport in graphene-based two-dimensional (2D) materials with an intrinsic superstructure, a secondary periodicity that matches the thermal phonon wavelength at room temperature. Coherent phonon transport in 2D materials is of particular interest due to its potential co-existence and interplay with other features of phonon transport, such as hydrodynamic phonon transport[14–16], enhanced resonant bonding[17] and electron-phonon interaction[18]. The two structures studied are metastable allotropes of graphene[19], named dodeca-graphene (D-graphene) and tetra-graphene (T-graphene)[20], as shown in Fig. 1(a)(b). We observe a drastic reduction, up to 80% and 73%, of the in-plane thermal conductivities of D-graphene and T-graphene compared to pristine graphene at room temperature despite similar bonding strengths, as shown in Fig. 1(c), which we attribute to the phonon-coherence-induced reduction of the phonon group velocity, enhancement of the phonon scatterings and suppression of hydrodynamic phonon transport. Our analysis provides quantitative understanding about the impact of phonon coherence on various aspects of thermal transport in 2D materials and demonstrates the possibility to tailor thermal conductivity of 2D materials without altering their fundamental compositions.

**Computational Methods**

The D-graphene unit cell contains 12 carbon atoms and is constructed by adjoining two graphene unit cells such that they do not share a side. The T-graphene has a square lattice with 4 carbon atoms in each unit cell. The crystal structures are shown in Fig. 1(a) and (b). Initial lattice parameters for D-graphene and T-graphene were taken from a previous study by Enyashin *et al.*[19] who reported a lattice constant of 6.83 Å and an average C-C distance of 1.449 Å for D-graphene and a lattice constant of 3.47 Å for T-graphene. The structures were constructed using these parameters and subsequently relaxed using density function theory (DFT) and the projector augmented wave method (PAW) as implemented in the Vienna *ab initio* Simulation Package (VASP)[21,22]. Calculations were performed using the Perdew-Burke-Ernzerhof (PBE) generalized gradient approximation (GGA) exchange correlation functionals. A $10 \times 10 \times 1$ Monkhorst-Pack k-mesh was used to sample the graphene, D-graphene, and T-graphene Brillouin Zones (BZ). A large vacuum space of 20 Å was introduced to minimize out of plane interactions that may have occurred through the use of periodic boundary conditions. All atomic positions were unconstrained and allowed to change until the maximum Hellman-Feynman forces were smaller than 0.01eV/Å per ion.

After relaxation, the initial C-C bonds in D-graphene degenerate into two distinct C-C bonds with lengths of 1.36 Å and 1.47 Å, and the lattice constant shrinks to 6.76 Å. The inter-graphene bonds are each 1.47 Å in length, with each alternating bond being 1.36 Å. The C-C bonds in T-graphene relax into identical bonds 1.44 Å long. The structures, shown in Fig 1(a) and (b), are metastable, with their total energy per atom being greater than that of pristine graphene, but existing in local energy minima.

For further DFT and thermal calculations, a $2 \times 2 \times 1$ D-graphene supercell was constructed, containing 48 atoms total. The supercell reveals a secondary symmetry, where six graphene unit cells adjoined by 1.47 Å bonds form a 12 sided carbon ring 5.27 Å in diameter. Likewise, a $4 \times 3 \times 1$ T-graphene supercell was constructed, also containing 48 atoms and revealing a secondary octagonal symmetry. Due to the low density of atoms at the unit cell corners, these superstructures result in highly porous structures, a design archetype that is commonly found in many engineered thermal metamaterials[9,10,23,24], but with a much smaller characteristic length scale. The second order, harmonic, interatomic force constants (IFCs) were obtained by utilizing density functional perturbation theory (DFPT) using the VASP package. Phonon dispersions for graphene, D-graphene, and T-graphene were calculated using these harmonic IFCs in concert with the PHONOPY package[25], and are shown in Fig. 2. Both D-graphene and T-graphene are dynamically stable as no imaginary frequencies are observed in their phonon dispersions.

The phonon Boltzmann transport equation (BTE) was solved using the ShengBTE[26] package in order to study the thermal transport properties of D-graphene. The lattice thermal conductivity tensor ($\kappa$) can be written as:

$$\kappa = \frac{1}{3}\sum_p \sum_k c_{ph} v_g(p,k)^2 \tau(p,k) ,$$

where p and q are the phonon branch and wave vector, respectively, $c_{ph}$ is the phonon specific heat capacity, $v_g = d\omega/dk$ is the phonon group velocity, and $\tau$ is the phonon lifetime and is also equivalent to the inverse of the phonon scattering rate. The calculation of $\kappa$ required the calculation of the third order, anharmonic IFCs. These were calculated using VASP, using the same supercell and k-mesh as for the 2nd order IFCs. Then, the phonon BTE was solved iteratively using ShengBTE. A $30 \times 30 \times 1$ grid was used for D-graphene while a $40 \times 40 \times 1$ grid was used for T-graphene. The thickness of D- and T-graphene was taken as 3.4 Å, the same as monolayer graphene.

**Results and Discussions**

The calculated lattice thermal conductivities of graphene, D-graphene and T-graphene are shown in Fig. 1(c). While the room temperature thermal conductivity of graphene is ~3000 W/mK, in good agreement with established literature[27], the room temperature thermal conductivities of D-graphene and T-graphene are ~600 W/mK and ~800 W/mK, 80% and 73% reduction from that of graphene, respectively. All thermal conductivities exhibit the expected 1/T trend and the discrepancy in thermal conductivity is greatly exacerbated at lower temperatures. This large reduction of thermal conductivity is unexpected. Firstly, the number density of carbon atoms in graphene, D-graphene and T-graphene are 0.383 Å$^{-2}$, 0.303 Å$^{-2}$ and 0.332 Å$^{-2}$, respectively. A first-order estimation using the effective medium theory[28], which considers the increased porosity of D-graphene and T-graphene but not the phonon-boundary scattering and coherence effects, would predict a thermal conductivity reduction of only 28% for D-graphene and 18% for T-graphene. Secondly, we examined the out-of-plane bonding strengths of the three carbon allotropes, which are responsible for the "stiffness" of the materials and the group velocity of the out-of-plane flexural phonons. Since the flexural phonons are major heat carriers in the three materials[27], their group velocity will have a strong influence on the thermal conductivity. The bonding strengths were quantified by calculating the change of the total potential energy of the allotropes when a carbon atom is displaced vertically (out-of-plane) from its equilibrium position. The quadratic coefficients of the energy change versus atomic displacement curves, as shown in Fig. 1(d), reflect the bonding strengths, whose numerical values are given in the inset table. The results indicate that D-graphene and T-graphene have similar out-of-plane bonding strengths, which are ~18% lower than that of graphene. As the acoustic phonon group velocity scales with the square root of the bonding strength (or the "spring constant"), this moderate reduction of the bonding strength still cannot explain the observed large difference in thermal conductivity between graphene and D- and T- graphene.

We hypothesize that the significant thermal conductivity reduction of D- and T-graphene is due to the phonon coherence effect, given that both structures contain a secondary periodicity compared to graphene. The D-graphene structure can be constructed by replacing each atom in a graphene honeycomb lattice with a hexagonal ring of carbon atoms. Similarly, the T-graphene structure can be constructed by combining a square lattice with a 4-atom square basis. As the periodicity of these superstructures (6.76 Å for D-graphene and 3.47 Å for T-graphene) is on the same order of the wavelength of the heat carrying acoustic phonons, it is expected that the interference of phonons with matching wavelengths will induce frequency gaps in the spectrum and suppress the phonon group velocity, as have been observed in artificial phononic crystals[4]. In the current case, however, these coherence effects are expected to have appreciable effects on the thermal conductivity even at room temperature due to the much smaller characteristic lengths.

To verity our hypothesis, we first analyze the phonon dispersions of the three allotropes, as shown in Fig. 2(a)-(c). All three materials exhibit the classic parabolic trend in its out of plane acoustic flexural (ZA) modes, typically found in 2D materials. More interestingly, The D-graphene and T-graphene dispersions show some markedly different behaviors from that of graphene. The maximum acoustic mode frequency in graphene is found in the longitudinal acoustic (LA) mode, and is ~40 THz. This is significantly depressed in D-graphene and T-graphene, with a maximum frequency of ~12.5 THz and ~18 THz, respectively, due to band folding. Furthermore, the coherent interference of zone-boundary phonons opens up frequency gaps in the dispersions in D-graphene and T-graphene. In the case of D-graphene, a frequency gap from ~1 THz to ~3 THz in the ZA branch is opened at the K point. In the case of T-graphene, a frequency gap around 8 THz in the ZA branch is opened at the X point. Although these are partial frequency gaps that do not cover the full BZ, they significantly suppress the group velocity of the ZA phonons and boost the phonon scatterings due to the flattened ZA dispersion and the increased ZA phonon density of states, as will be discussed later in detail.

Another interesting phenomenon is the occurrence of three low-lying optical (LLO) phonons modes in D-graphene that cross and interact heavily with the acoustic modes. Branch 4, the lowest lying optical mode and referred to as the ZO mode, couples heavily with the ZA mode along the BZ boundary. Branch 5 interacts strongly with the TA mode. Branch 5 and the LA mode are coupled heavily along the M-K path. Branch 6 exhibits four distinct crossing events with the LA mode in the irreducible wedge. These LLO modes arise from the folding of the acoustic modes, which can be clearly seen along the Γ-M path. This effect is most clearly seen in the ZA and TA branches, with the associated folded acoustic modes being modes 4 and 5, respectively. Physically, the LLO phonons in D-graphene represent the coherent collective motions of the six carbon atoms in hexagonal ring units, or "meta-atoms". As shown in Fig. 2(d), the ZO branch represents the bending of the unit cell about its center, with atoms at the edge of the unit cell being translated upwards out of plane while an equal and opposite translation occurs in atoms near the unit cell center. Branch 5 resembles the twisting of the unit cell. Branch 6 represents whole-scale, out of phase, out of plane translations of the individual graphene unit cells. These LLO phonons are unique signatures originated from the secondary periodicity of D-graphene and are expected to strongly scatter the acoustic phonons and further suppress the thermal conductivity of D-graphene.

We further examine the phonon-phonon scattering rates of the acoustic modes in the three carbon allotropes, as shown in Fig. 3. The scattering rates of the ZA modes, shown in Fig. 3(a), in all three allotropes show similar scaling with the phonon frequency. The scattering rates in D-graphene and T-graphene are higher than those in graphene, mainly due to the increased phonon density of states in this frequency region, as shown in Fig. 3(d), which in turn is caused by the suppressed group velocity of the acoustic phonons in D-graphene and T-graphene. Of particular interest is the enhancement of phonon scattering near the frequency gaps, which is expected to happen due to the local high phonon density of states and enlarged phonon scattering phase space. This effect is clearly exhibited in the scattering rates of TA and LA phonons, as shown in Fig. 3(b) and (c). Clear peaks of scattering rates of both TA and LA phonons in D-graphene appear near 3 THz, which coincides with the upper edge of the frequency gap in D-graphene, as marked by the arrows in Fig. 3(b) and (c). In the case of T-graphene, upturns of the scattering rates for both TA and LA phonons are observed near 10 THz, as a result of strong hybridization between the ZA mode and the low-lying optical mode. These distinct features will significantly contribute to the reduction of the thermal conductivity of D-graphene and T-graphene.

In Table I, we analyze the mode-specific contributions to the thermal conductivity. Acoustic modes contribute to almost the entirety of the conductivity of graphene and the out of plane acoustic (ZA) contribution comprises the majority of the conductivity, contributing 83.85%, which agrees with values previously reported[27]. Remarkably, the ZA contribution in D-graphene and T-graphene is reduced significantly, falling to roughly 47% and 60%, respectively. In addition to the factors discussed previously, namely the reduced group velocity and enhanced scattering rates, another important factor is the breakdown of the hydrodynamic transport of the ZA phonons in D-graphene and T-graphene[14,16]. In graphene, the quadratic dispersion and strong anharmonicity of the ZA phonons lead to dominant momentum-conserving normal phonon-phonon scattering over the momentum-destroying Umklapp phonon-phonon scattering, which gives rise to hydrodynamic transport of the ZA phonons[14]. In the hydrodynamic regime, the phonon transport is much less dissipative, an effect that significantly contributes to the extraordinary high thermal conductivity of graphene. In D-graphene and T-graphene, however, the hydrodynamic transport of ZA phonons is expected to be suppressed, due to the shrunk Brillouin zone and the flattened ZA phonon dispersions. We verify this hypothesis by explicitly calculating the normal and Umklapp phonon-phonon scattering rates in all three carbon allotropes at 100 K and 300 K, as displayed in Fig. 4. In the case of graphene, most clearly seen at a lower temperature (100 K, Fig. 4a), the normal phonon scattering rates are one to two orders of magnitude higher than the Umklapp phonon scattering rates. Even at room temperature (300 K, Fig. 4d), the normal phonon scatterings still play the dominate role. In the case of T-graphene and D-graphene, however, the normal and Umklapp phonon scattering rates are at comparable levels at both 100 K and 300 K, especially for the low-frequency acoustic phonons which are the major heat carriers in these materials. The breakdown of hydrodynamic phonon transport in D-graphene and T-graphene will significantly increase the dissipation during heat conduction and further suppress their thermal conductivities.

In summary, we apply first-principles phonon simulation to understand the thermal transport of carbon allotropes D-graphene and T-graphene with a secondary periodicity compared to graphene. Although D-graphene and T-graphene are 2D carbon materials constructed out of graphene unit cells and similar carbon-carbon bonds and thus possess a fundamental composition favorable to efficient thermal transport, we find the thermal conductivities of D-graphene and T-graphene are significantly reduced from that of graphene. The introduction of a secondary in-plane periodicity results in the folding of the acoustic mode dispersions in D-graphene and T-graphene, creating frequency gaps at the Brillouin zone boundaries due to coherent phonon interference. This effect leads to suppressed ZA phonon group velocity, enhanced phonon scattering near the frequency gaps and the breakdown of hydrodynamic phonon transport, reducing the overall lattice thermal conductivity. The four to five times reduction in thermal conductivity is attributed almost entirely to the introduction of the secondary periodicity of the superstructure, as the main structure inherent to graphene and the carbon-carbon bond strengths have largely been maintained. Our study presents D-graphene and T-graphene as ideal platforms to explore the effect of phonon coherence on thermal conductivity at practical temperatures and provides quantitative insights about manipulating heat conduction utilizing the wave nature of thermal phonons.

**Acknowledgements**

This work is based on research supported by the U.S. Department of Energy, Office of Basic Energy Sciences, Division of Materials Science and Engineering through the Early Career

Research Program under the award number DE-SC0019244. B. L. acknowledges the support provided by the Regents' Junior Faculty Fellowship and an Academic Senate Faculty Research Grant from UCSB. We acknowledge computational resource support from the Center for Scientific Computing from the CNSI, MRL: an NSF MRSEC (DMR-1720256) and NSF CNS-1725797. This work also used the Extreme Science and Engineering Discovery Environment (XSEDE)[29] Stampede 2 at the Texas Advanced Computing Center (TACC) through allocation TG-DMR180044. XSEDE is supported by NSF grant number ACI-1548562.

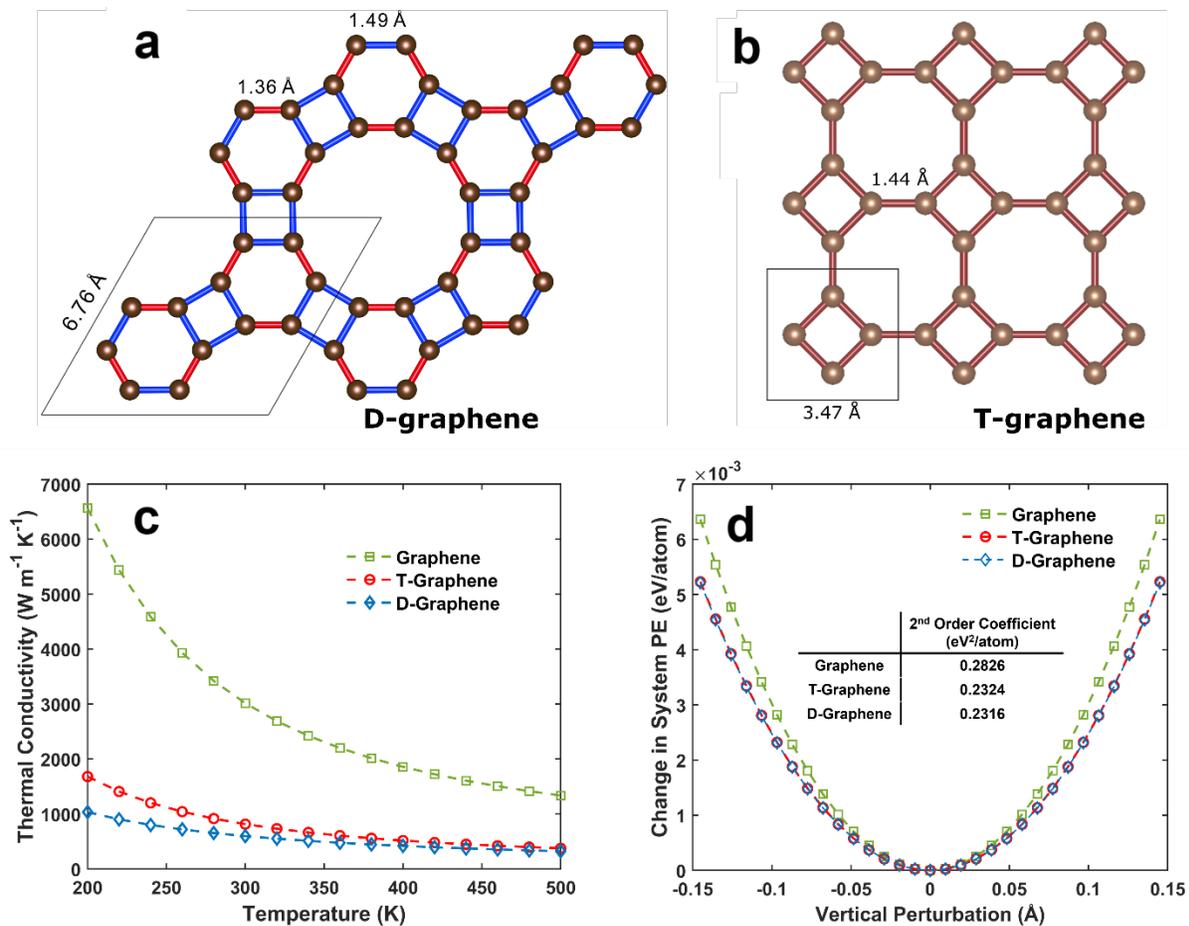

**Figure 1.** Crystal structure and lattice parameters of (a) D-graphene and (b) T-graphene. (c) Calculated thermal conductivities of graphene, T-graphene and D-graphene versus temperature. (d) Changes in total potential energy versus vertical perturbation of a carbon atom in graphene, T-graphene and D-graphene, indicating the bonding strengths in the three carbon allotropes. The quadratic coefficients of the curves are quantitative descriptors of the bonding strengths and are given in the inset table. PE: potential energy.

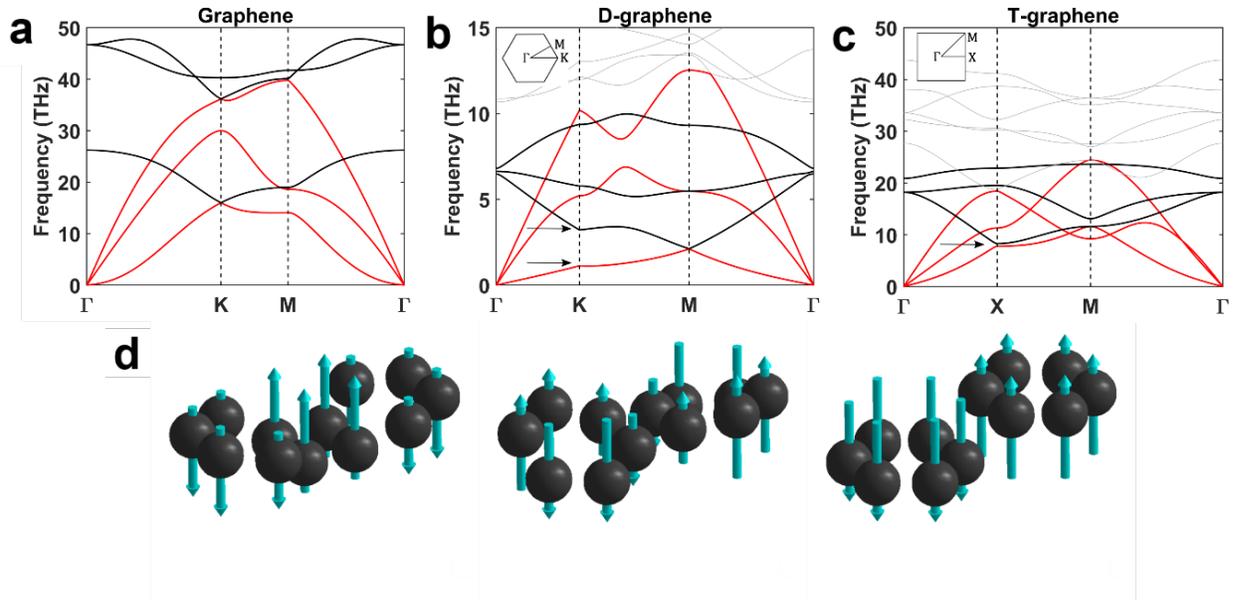

**Figure 2.** Calculated phonon dispersion relations for (a) graphene, (b) D-graphene and (c) T-graphene. The arrows in (b) and (c) indicate the frequency gaps opened due to coherent phonon interference. (d) Vibrational modes of the three low lying optical phonons in D-graphene, indicating their origins as collective motions of base units.

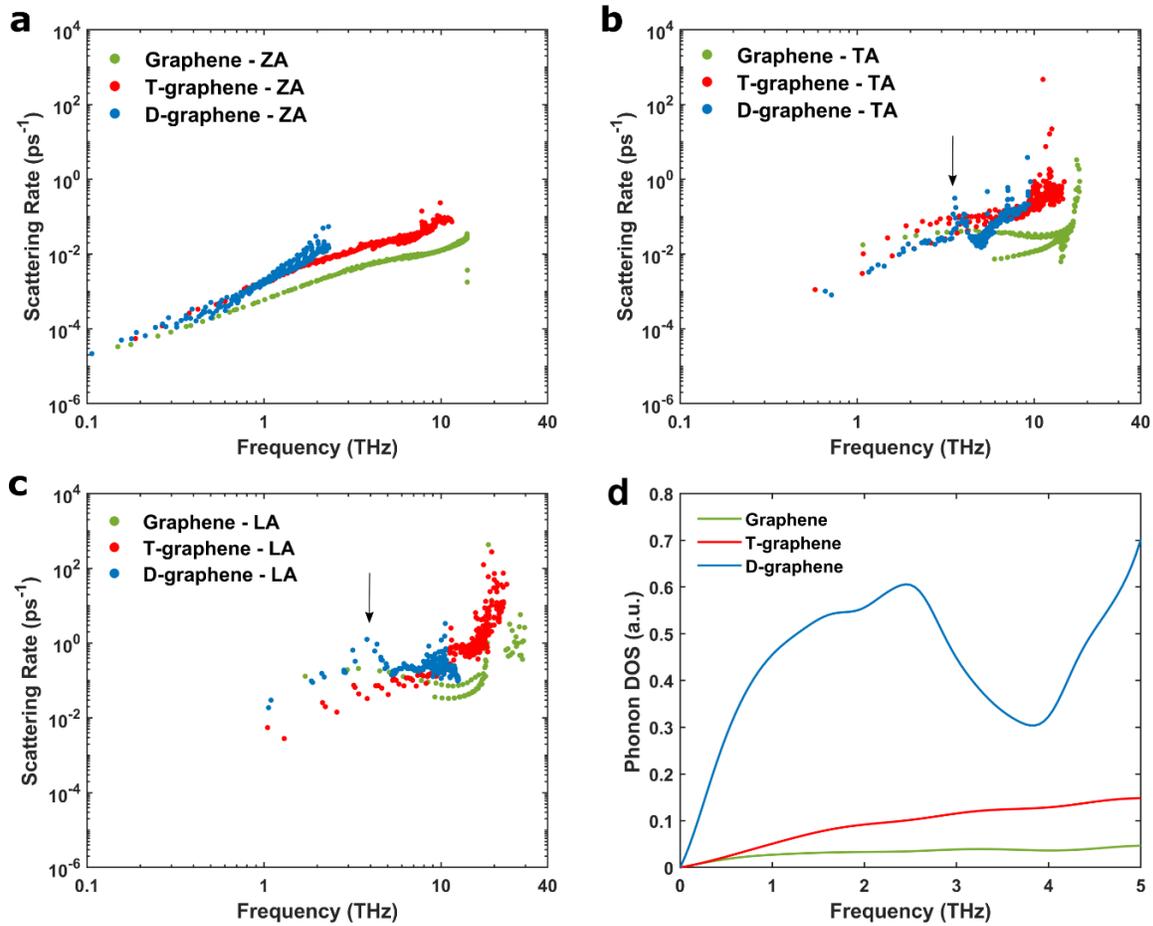

**Figure 3.** Calculated phonon-phonon scattering rates of (a) ZA branches, (b) TA branches and (c) LA branches in graphene, T-graphene and D-graphene. The arrows in (b) and (c) indicate the peaks in the scattering rates for TA and LA branches in D-graphene due to the frequency gap of the ZA branch. (d) Phonon density of states (DOS) of the three allotropes versus phonon frequency (0 to 5 THz).

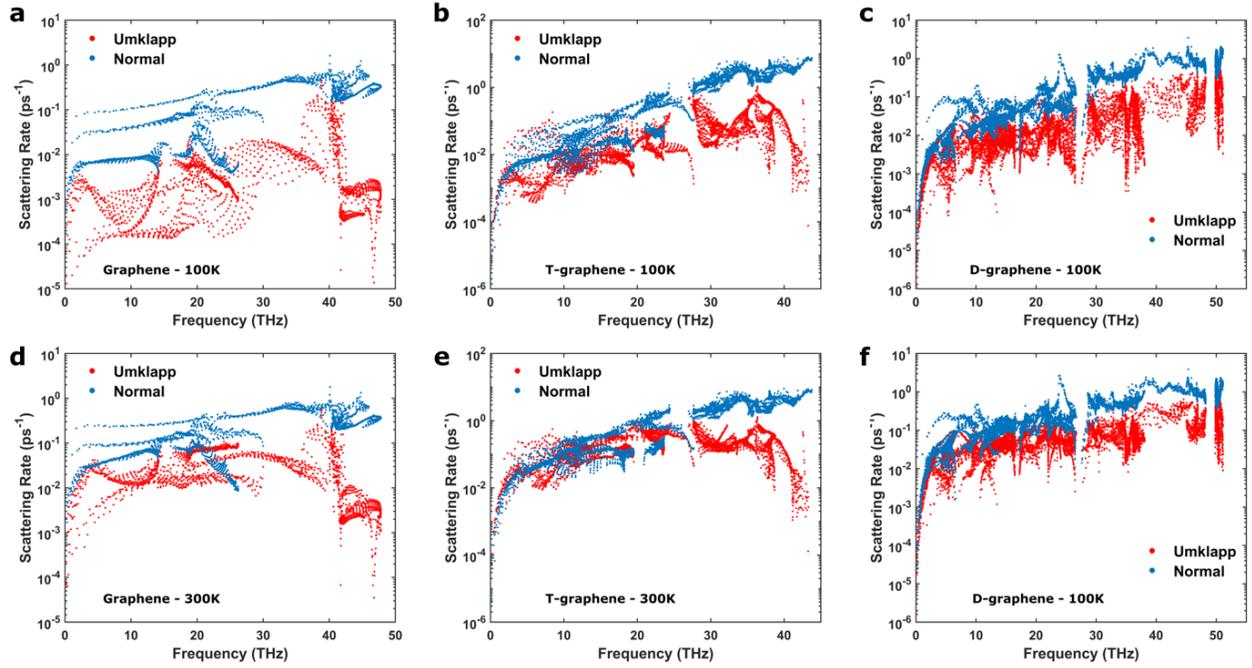

**Figure 4.** Calculated phonon-phonon normal and Umklapp scattering rates of the three carbon allotropes at 100 K (a)-(c), and 300 K (d)-(f), indicating that hydrodynamic phonon transport in graphene is largely suppressed in T-graphene and D-graphene due to comparable levels of normal and Umklapp scatterings, leading to significant reduction of their thermal conductivities compared to that of graphene.

**Table I** Mode-specific contribution to the thermal conductivity in three carbon allotropes.

|            | Acoustic Contribution | Optical Contribution | ZA Contribution |
|------------|----------------------|---------------------|-----------------|
| Graphene   | 97.14%               | 2.86%               | 83.85%          |
| T-graphene | 90.62%               | 9.38%               | 60.17%          |
| D-graphene | 81.85%               | 18.15%              | 46.83%          |